# Application of Real-time Digitization Technique in Beam Measurement for Accelerators

ZHAO Lei(赵雷) [1,2] ZHAN Lin-song (占林松) [1,2]
GAO Xing-shun (高兴顺) [1,2] LIU Shu-Bin(刘树彬) [1,2] AN Qi(安琪) [1,2;1)]
1 State Key Laboratory of Particle Detection and Electronics, University of Science and Technology of China, Hefei, 230026, China,
2 Department of Modern Physics, University of Science and Technology of China, Hefei, 230026, China

**Abstract:** Beam measurement is very important for accelerators. With the development of analog-to-digital conversion techniques, digital beam measurement becomes a research hot spot. IQ (In-phase & Quadrature-phase) analysis based method is an important beam measurement approach, the principle of which is presented and discussed in this paper. The State Key Laboratory of Particle Detection and Electronics in University of Science and Technology of China has devoted efforts to the research of digital beam measurement based on high-speed high-resolution analog-to-digital conversion, and a series of beam measurement instruments were designed for China Spallation Neutron Source (CSNS), Shanghai Synchrotron Radiation Facility (SSRF), and Accelerator Driven Sub-critical system (ADS).

**Key words:** digital beam measurement; beam phase and position; high-speed analog-to-digital conversion; digital signal processing

**PACS:** 84.30.-r, 07.05.Hd

## 1. Introduction

Accelerators are widely applied in scientific research and many other domains. In China, great efforts are devoted to accelerator-based facilities, such as China Spallation Neutron Source (CSNS), Accelerator Driven Sub-critical system (ADS), Beijing Electron Positron Collider (BEPC), Shanghai Synchrotron Radiation Facility (SSRF), National Synchrotron Radiation Laboratory (NSRL) and Heavy Ion Research Facility in Lanzhou (HIRFL).

High quality beam measurement is very important for accelerators, and thus is a worldwide research hotspot [1-19]. A series of beam measurement techniques were developed for different types of accelerators; however, with higher requirement of beam measurement and development of electronics, more advanced measurement techniques need to be researched for more precise diagnostic of beam parameters.

Traditional beam measurement techniques are based on analog signal manipulation, for example, in Advanced Light Source in USA, Pohang Light Source (PLS) in South Korea, and BEPCII etc. [1-6]. With the development of high-speed high-resolution Analog-to-Digital (A/D) conversion and digital signal processing, digital beam measurement becomes possible. Compared with the analog method, higher precision and flexibility can be achieved. Researchers in many institutes focus on the research in this domain, such as in Low Energy Demonstrator Accelerator (LEDA) [7, 9], Spallation Neutron Source (SNS) in USA [10], Gesellschaft für Schwerionenforschung (GSI) in Germany [11], Institute of High Energy (IHEP), CAS, NRSL [14, 15], SSRF [12, 13, 17]. There are also companies specializing in beam measurement instrument design. For instance, the Libera Electron and Single Pass H series [18, 19] from Instrumentation Technologies (IT) Company in Slovenia are widely employed in accelerators.

As one of the beam measurement research groups, the State Key Laboratory of Nuclear Detection and Electronics also made efforts in this domain, and focused on applying real-time digitization in beam phase (energy)

*Supported by Knowledge Innovation Program of the Chinese Academy of Sciences (KJCX2-YW-N27), National Natural Science Foundation of China (11205153, 10875119), the Fundamental Research Funds for the Central Universities (WK2030040029), and the CAS Center for Excellence in Particle Physics (CCEPP).
1) Email: anqi@ustc.edu.cn

and position measurement.

This paper is organized as follows. Modern beam phase and position measurement methods based on IQ (In-phase and Quadrature-phase) analysis are discussed, and then three beam measurement systems designed in our research are presented, including the system structure, kernel techniques and performance.

## 2. Principle of beam phase and position measurement method based on IQ analysis

Beam measurement methods based on IQ analysis are widely used in accelerators [2-7], [9-13], [17-19]. The basic idea is to obtain the I and Q values of the beam Radio Frequency (RF) signals, and then use them for beam measurement.

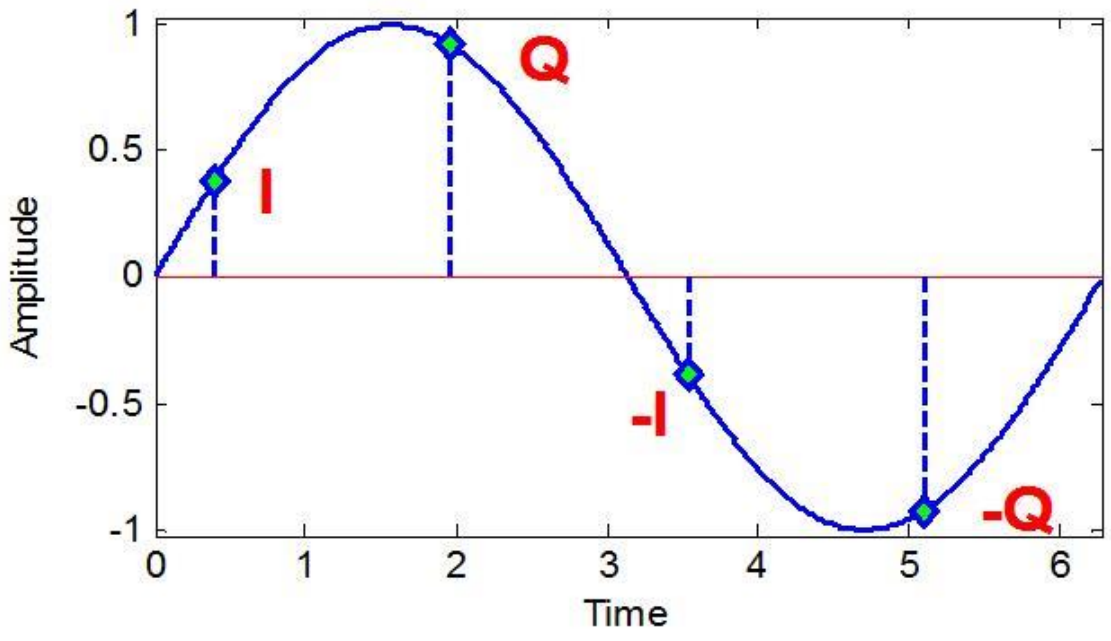

Fig. 1 Principle of the beam measurement based on IQ analysis method.

As shown in Fig. 1, by filtering the beam RF signal, a sinusoid signal can be obtained, which corresponds to the component of a certain order of harmonic in the RF signal; then if we can obtain the I and Q values, the phase information of the beam can calculated as:

$$\theta = \arctan\left(\frac{I}{Q}\right) \quad (1).$$

Meanwhile, the signal amplitude can also be calculated as:

$$Amplitude = \sqrt{I^2 + Q^2} \quad (2).$$

Special pickups will be used for beam position measurement. For example, in the Storage Ring in SSRF, Beam Position Monitor (BPM) pickups are employed. As shown in Fig. 2, when the beam passes through the cross section of the BPM, four signals are generated through A, B, C, and D. By calculating the amplitudes of these four signals as in (2), beam position can be finally obtained according to the Δ/Σ algorithm, as in

$$x = K_x \frac{(V_A + V_D) - (V_B + V_C)}{V_A + V_B + V_C + V_D} \quad (3)$$

$$y = K_y \frac{(V_A + V_B) - (V_C + V_D)}{V_A + V_B + V_C + V_D} \quad (4)$$

where $V_A$, $V_B$, $V_C$, and $V_D$ refer to the amplitudes of four signals; $K_X$ and $K_Y$ are the effective length factors in X and Y directions.

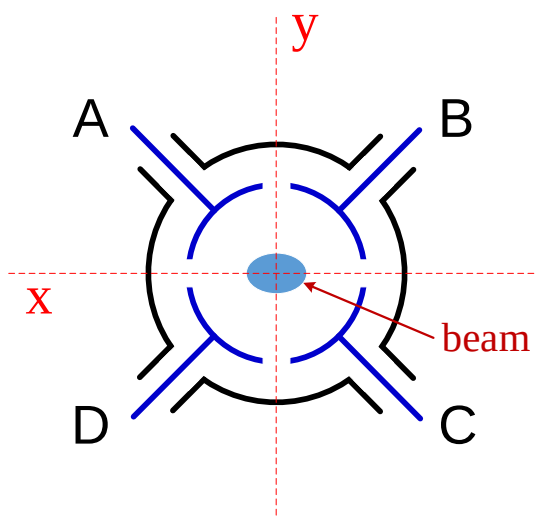

Fig. 2 Basic structure of BPM.

The kernel task in IQ analysis is to obtain the I and Q value, the approach for which can be categorized into two main techniques – IQ demodulation and IQ sampling. In both ways, the beam RF signals first pass through Band Pass Filters (BPFs) to extract the fundamental or higher order harmonic, and are then processed by different methods.

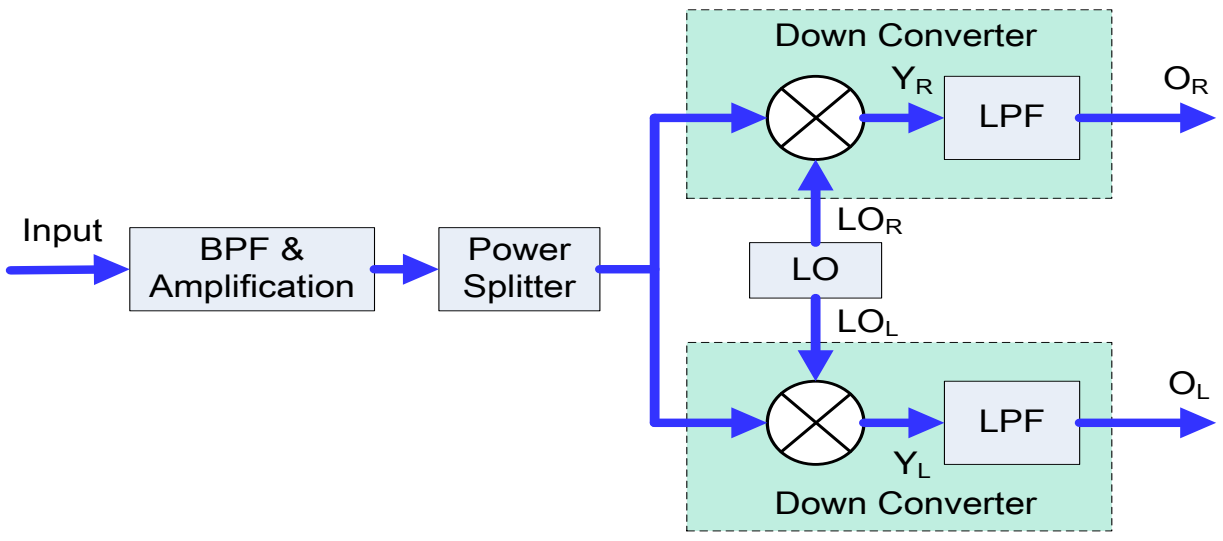

Fig. 3 Block diagram of the IQ demodulation method.

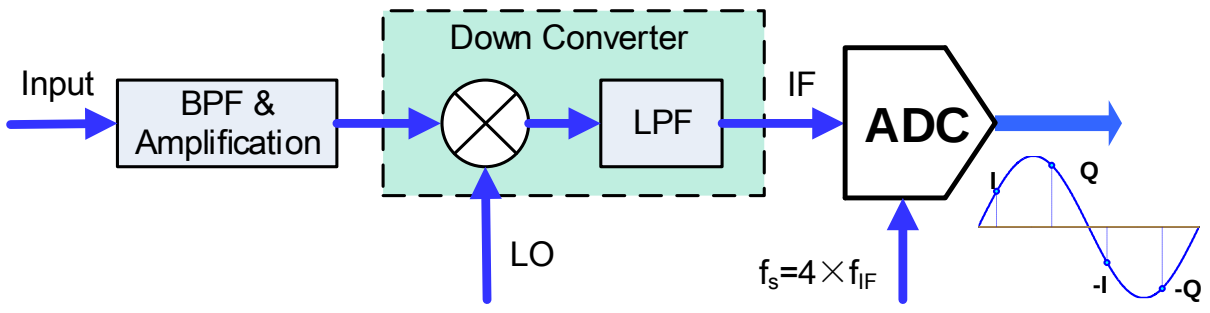

Fig. 4 Block diagram of the IQ sampling method.

Shown in Fig. 3 are the block diagram of the IQ demodulation method. A sinusoid signal is generated after BPFs and amplifiers, and then split into two paths, which are mixed with two orthogonal Local Oscillator (LO) signals. The outputs of the mixers can be expressed as

$$\begin{aligned} Y_R &= Signal_{IN} \times LO_R = A\sin(2\pi f_{IN}t + \varphi)\cdot\sin(2\pi f_{LO}t) \\ &= \frac{1}{2}A\left\{\cos\left[2\pi(f_{IN} - f_{LO})t_s + \varphi\right] - \cos\left[2\pi(f_{IN} + f_{LO})t_s + \varphi\right]\right\} \\ Y_L &= Signal_{IN} \times LO_L = A\sin(2\pi f_{IN}t + \varphi)\cdot\cos(2\pi f_{LO}t) \\ &= \frac{1}{2}A\left\{\sin\left[2\pi(f_{IN} - f_{LO})t_s + \varphi\right] - \sin\left[2\pi(f_{IN} + f_{LO})t_s + \varphi\right]\right\} \end{aligned} \quad (5).$$

(5) indicates that the outputs contain the difference and sum frequency components of the input RF signal frequency ($f_{IN}$) and the LO frequency ($f_{LO}$). Passing through Low Pass Filters (LPFs), the two outputs are converted to

$$\begin{aligned} O_R &= \frac{1}{2}A\cos\left[2\pi(f_{IN} - f_{LO})t_s + \varphi\right] \xrightarrow{f_{IN}=f_{LO}} \frac{1}{2}A\cos(\varphi) \\ O_L &= \frac{1}{2}A\sin\left[2\pi(f_{IN} - f_{LO})t_s + \varphi\right] \xrightarrow{f_{IN}=f_{LO}} \frac{1}{2}A\sin(\varphi) \end{aligned} \quad (6),$$

which indicates that the phase ($\varphi$) and amplitude ($A$) information of the original RF signal are contained in the two outputs $Q_R$ and $Q_L$. When $f_{LO}$ is set to be equal to $f_{IN}$, only DC components remains, which corresponds to the I and Q values. Traditional methods to implement this technique are based on complete analog signal manipulation; nowadays, part of or even complete signal processing can be performed in digital signal processing domain.

As for the other way, IQ sampling, as shown in Fig.4, the sinusoid signal is first down converted to an IF signal. Since down conversion actually consists of the mixer and LPF as in (5) and (6), the IF signal also contains the phase and amplitude information of the original RF signal. Then this IF signal is digitized with a sampling frequency of just four times of the IF signal frequency, so there exist exactly four samples in each period of the digitized IF signal, which are I, Q, -I, and –Q values. With this method, the I and Q values are obtained through sampling, thus the analog circuits complexity is reduced.

We implemented digital beam measurement based on basic idea of the IQ analysis method and designed three beam measurement systems, which will be presented in the following sections.

## 3. Fully Digital Beam Position Measurement System in SSRF

SSRF is one of the third-generation light sources in the world. In its Storage Ring, up to 720 electron bunches circulate with a duty ratio of 500:220 and a Turn-By-Turn (TBT) frequency of $f_{mc}$=693.964 kHz. Signals from the BPM pickups are high frequency pulses with a repetition frequency of 499.654 MHz ($f_{mc}$×720). The waveform of single pulse and the frequency spectrum of the pulse sequence are shown in Fig. 5 [13] and Fig. 6, respectively.

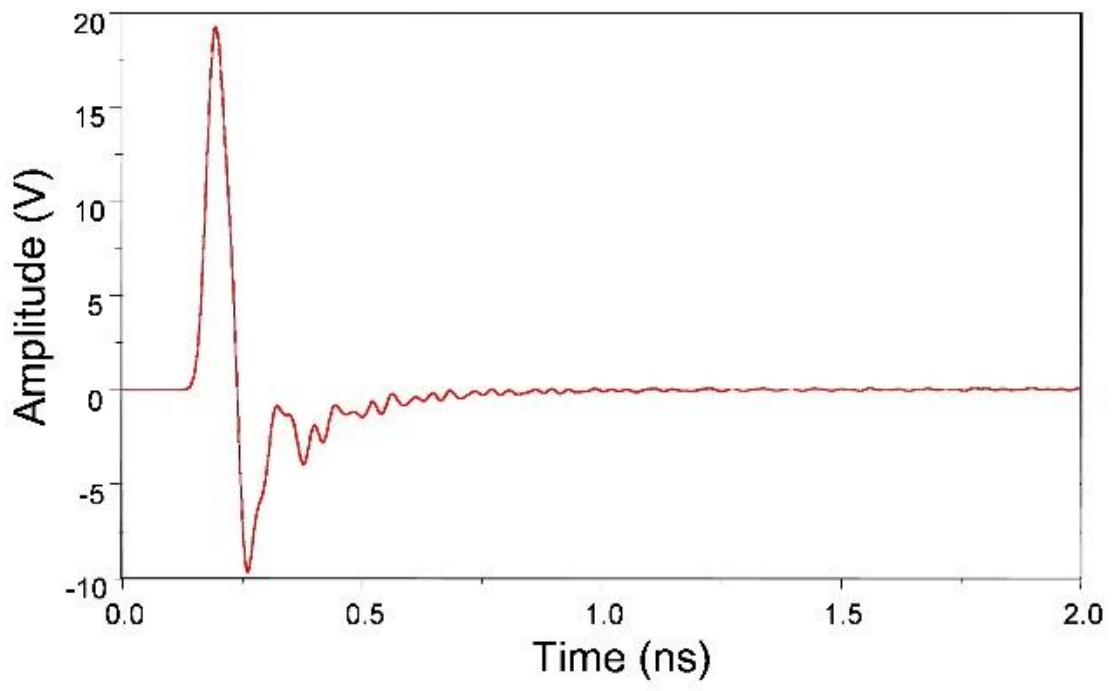

Fig. 5 Waveform of the beam signal.

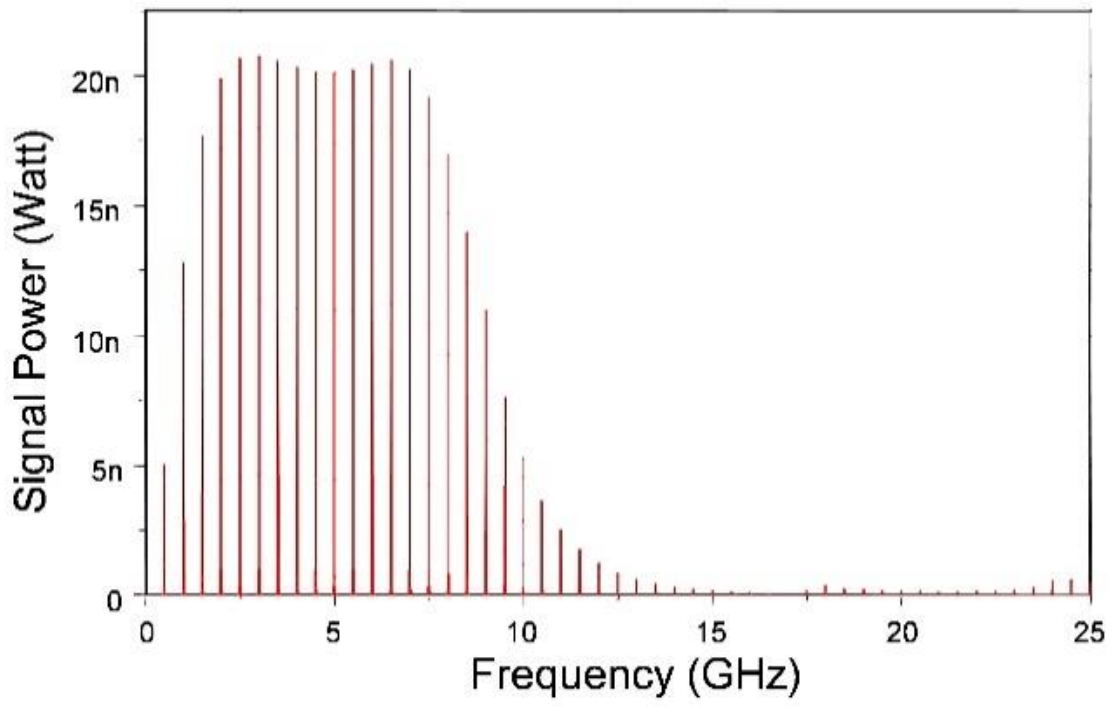

Fig. 6 Frequency spectrum of the beam signal.

Shown in Fig. 7 is the structure of the beam position measurement electronics [20-23]. After amplification and band pass filtering, the RF signal is directly digitized by high-speed ADCs with a sampling frequency of 117.2799 MHz, through which a digital IF signal of 30.5344 MHz is obtained. Based on the IQ demodulation method, this IF signal is moved to DC (i.e. I and Q values are obtained) through Digital Down Conversion (DDC) algorithm in the FPGA, as shown in Fig. 8. Then the amplitudes of four channels can be calculated, with which the position information can be finally obtained.

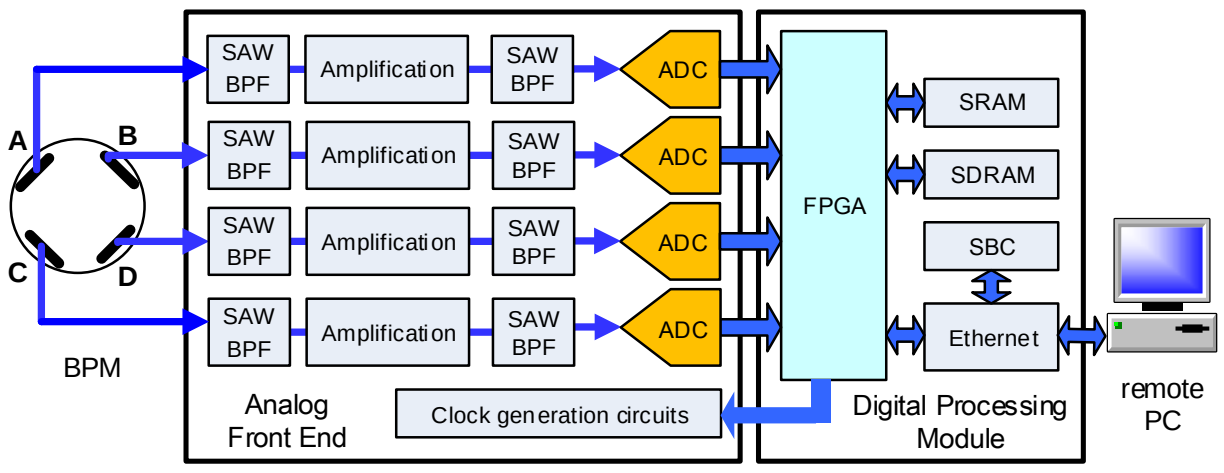

Fig. 7 Block diagram of the fully digital beam position measurement system.

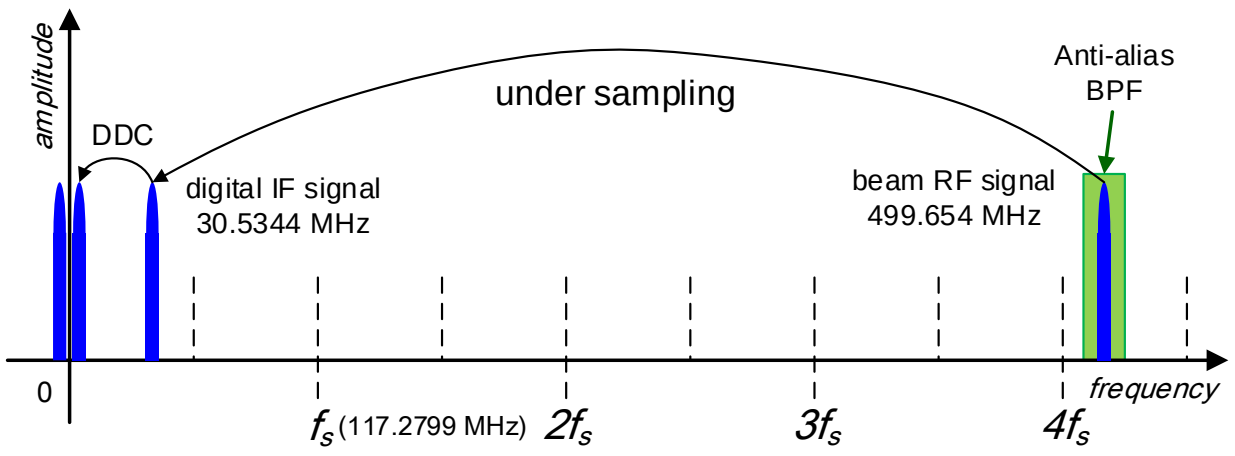

Fig. 8 Signal processing in frequency domain.

The structure of the kernel digital signal processing algorithm is shown in Fig. 9. The digital IF signal is mixed with the two orthogonal outputs from a Numerically Controlled Oscillator (NCO) to obtain the I and Q values, which are then processed by LPF (consisting of CIC filter and FIR filter) to enhance the Signal-to-Noise Ratio (SNR) of the signal while decreasing the data rate from 117.2799 MHz to the TBT rate (693.964 kHz). Through the CORDIC logic, the amplitudes can be calculated with the I and Q values, and then beam position can be finally obtained based on the $\Delta/\Sigma$ algorithm.

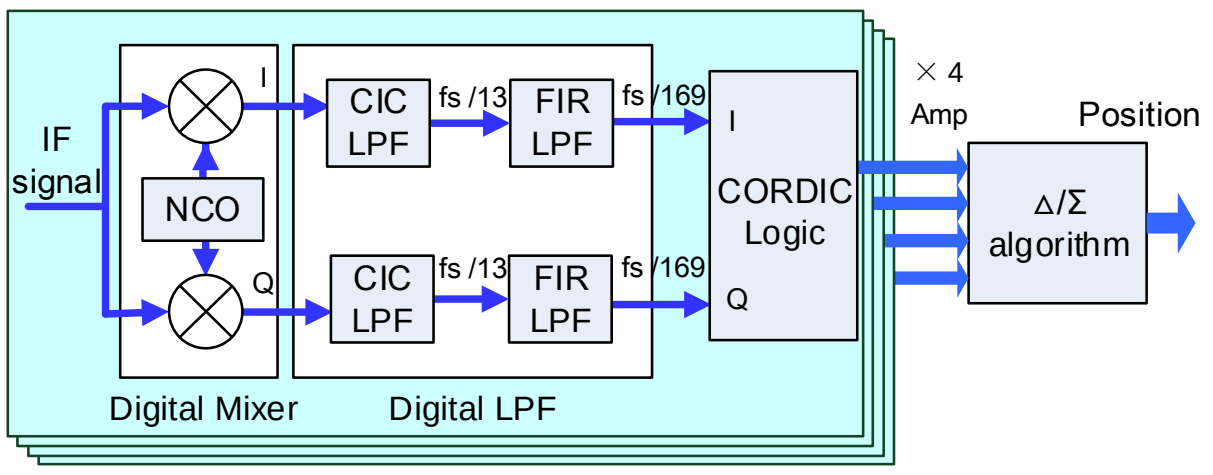

Fig. 9 Block diagram of the kernel digital signal processing algorithm.

A series of test were conducted to evaluate the system performance both in the laboratory and with the beam in SSRF. Shown in Fig. 10 is the beam position measurement resolution test results, which indicate that the TBT position resolution is better than 10 μm ($K_X$=$K_Y$=10 mm) within the input signal amplitude range from -40 dBm to 10 dBm.

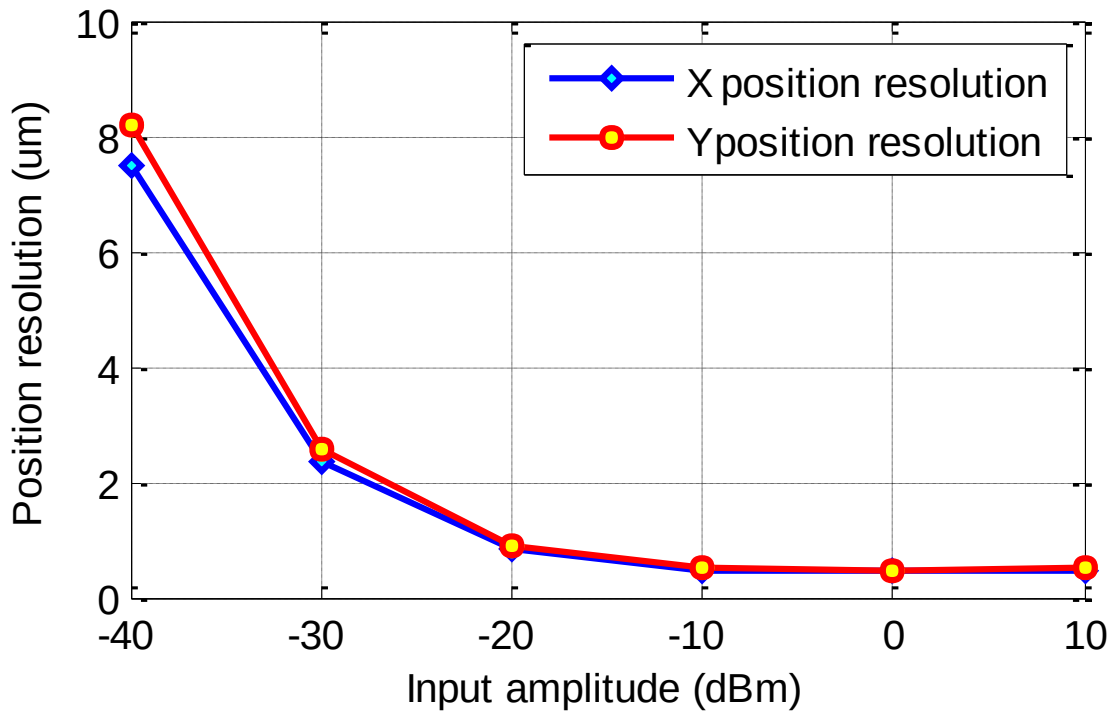

Fig. 10 Position resolution in X and Y directions.

We also conducted tests on the Libera series of digital beam measurement instruments for performance comparison, which are widely applied in accelerators in the world. Fig. 11 and Fig. 12 are the test results of Libera Electron and its new version Brilliance.

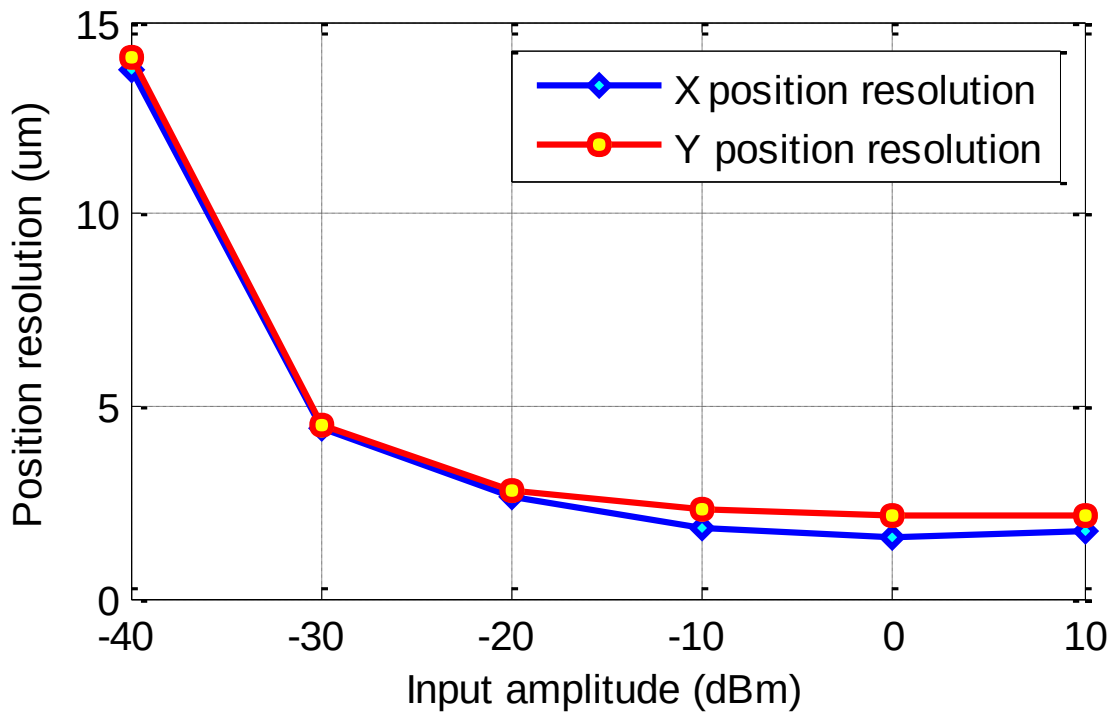

Fig. 11 Position resolution in X and Y directions of Libera Electron.

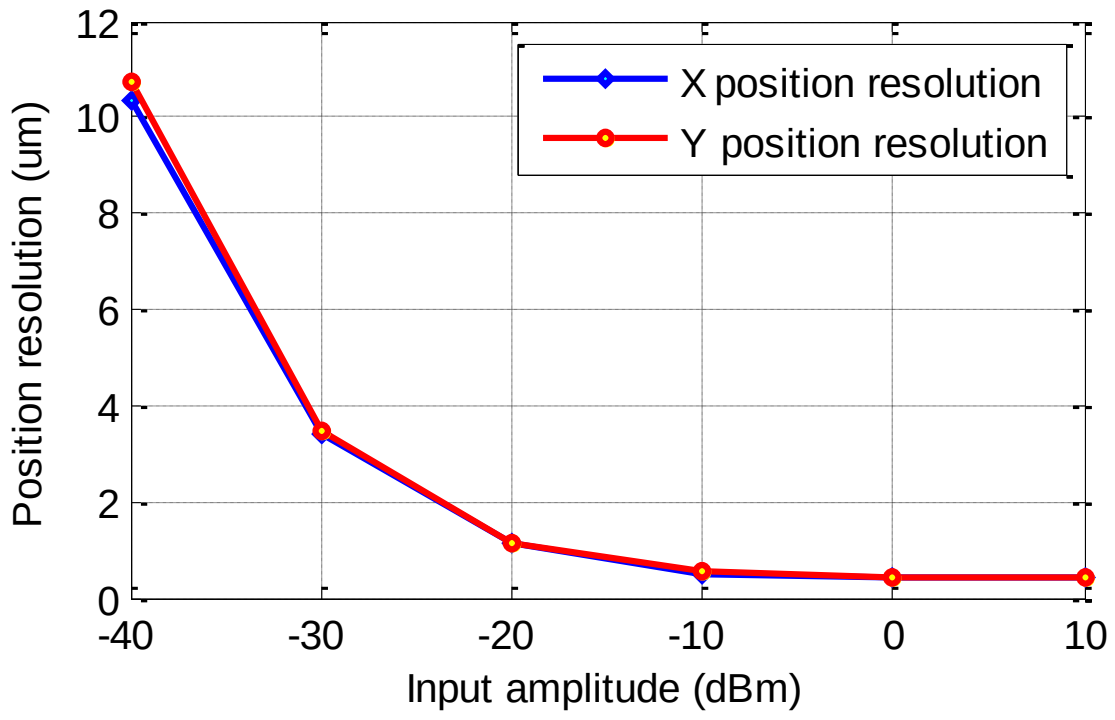

Fig. 12 Position resolution in X and Y directions of Libera Brilliance.

Shown in Fig. 13 is the beam position measurement system under commissioning tests in SSRF.

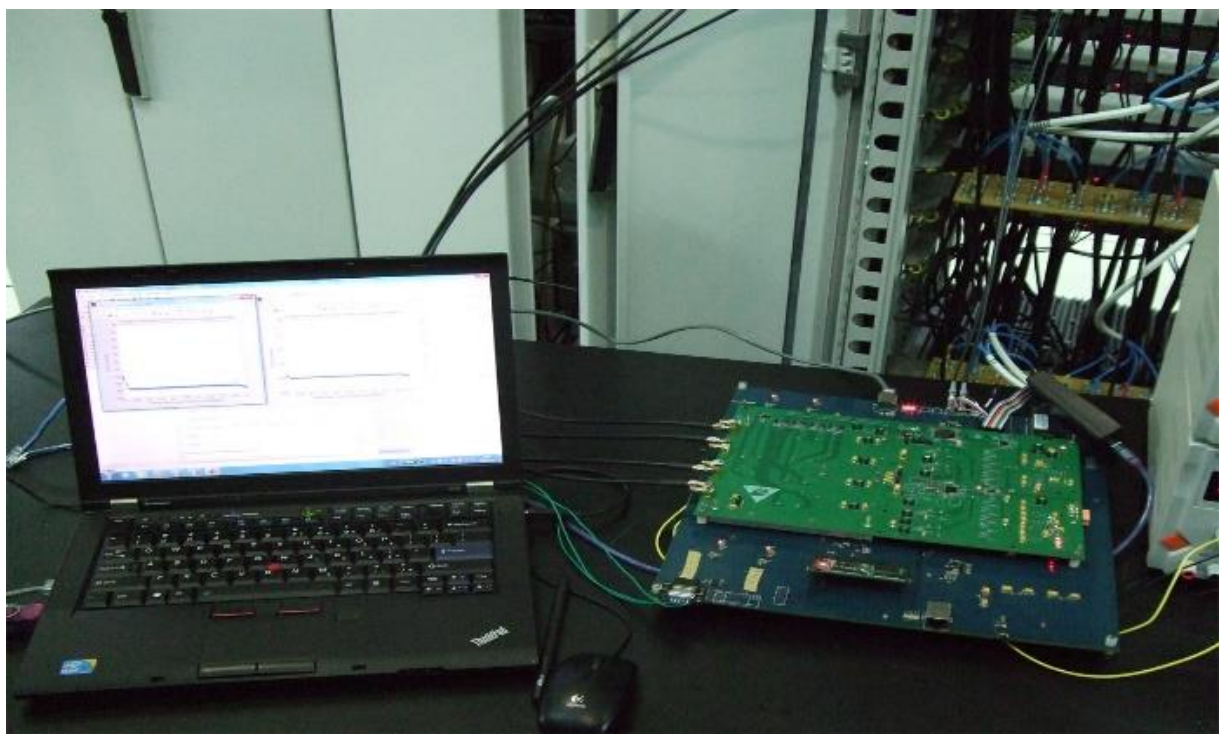

Fig. 13 Setup of the commissioning tests in SSRF.

Fig. 14 and Fig. 15 show the histogram of the Y position measurement results and its frequency spectrum, which correspond to a position resolution of 0.67 μm, and the frequency components close to DC concord well with the beam behavior.

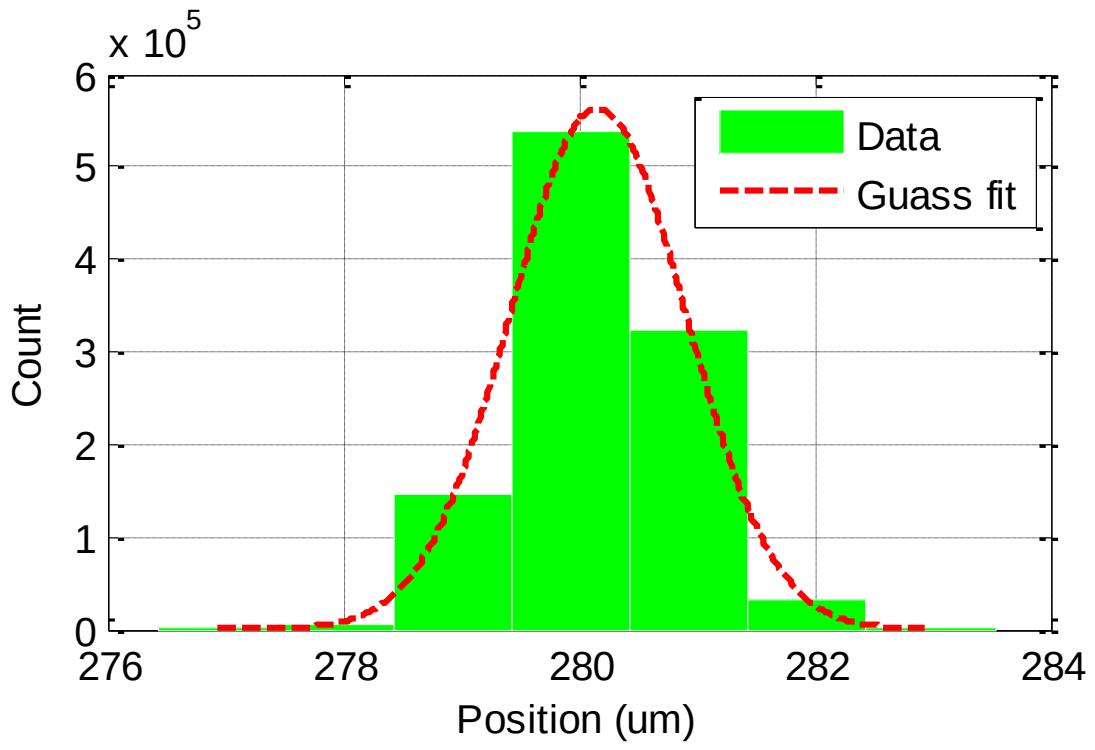

Fig. 14 Histogram of the Y position measurement results.

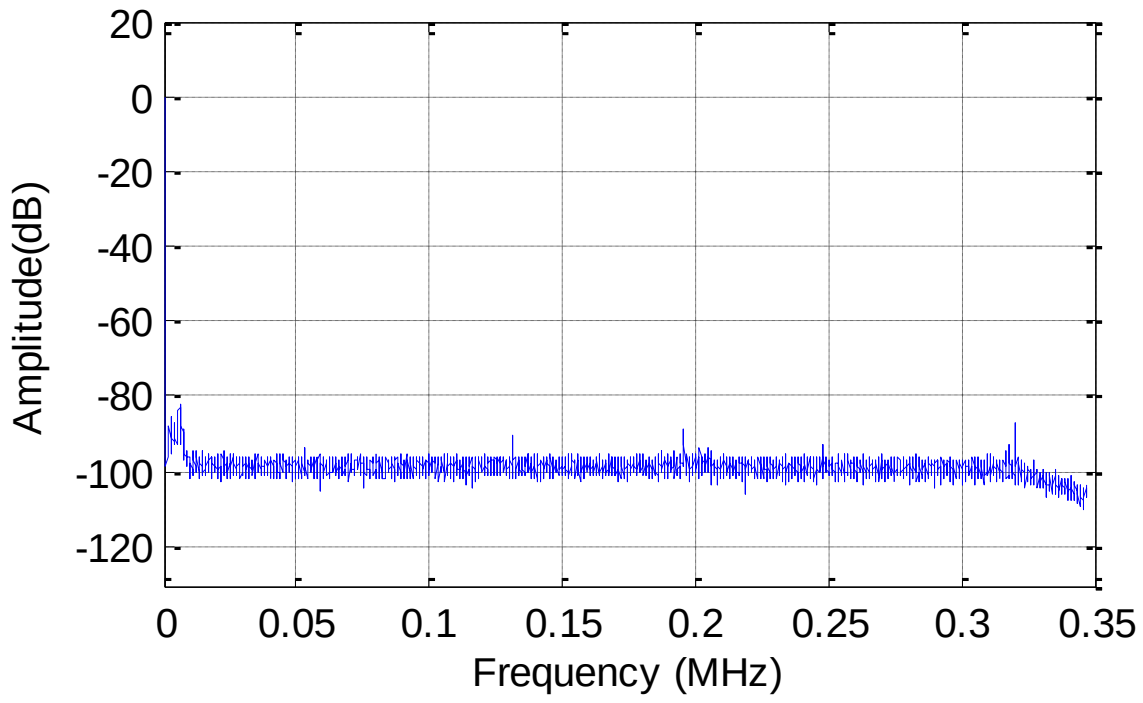

Fig. 15 Frequency Spectrum of the Y position measurement results.

We also conducted tests during the beam injection process. The results are shown in Fig. 16, in which the fluctuation can be clearly observed, well as expected. Fig. 17 shows the normalized frequency spectrum of the beam position measurement results of our system and the Libera Brilliance instrument, which indicate the noise floor is almost equivalent.

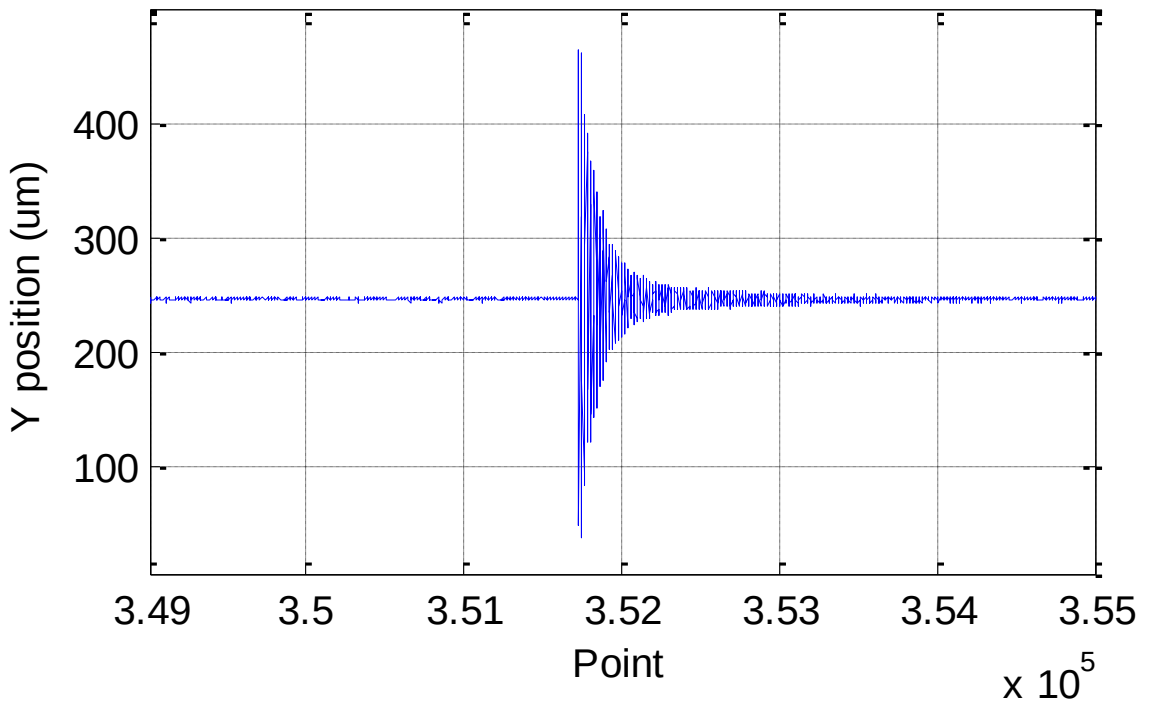

Fig. 16 Waveform of the Y position measurement results during injection.

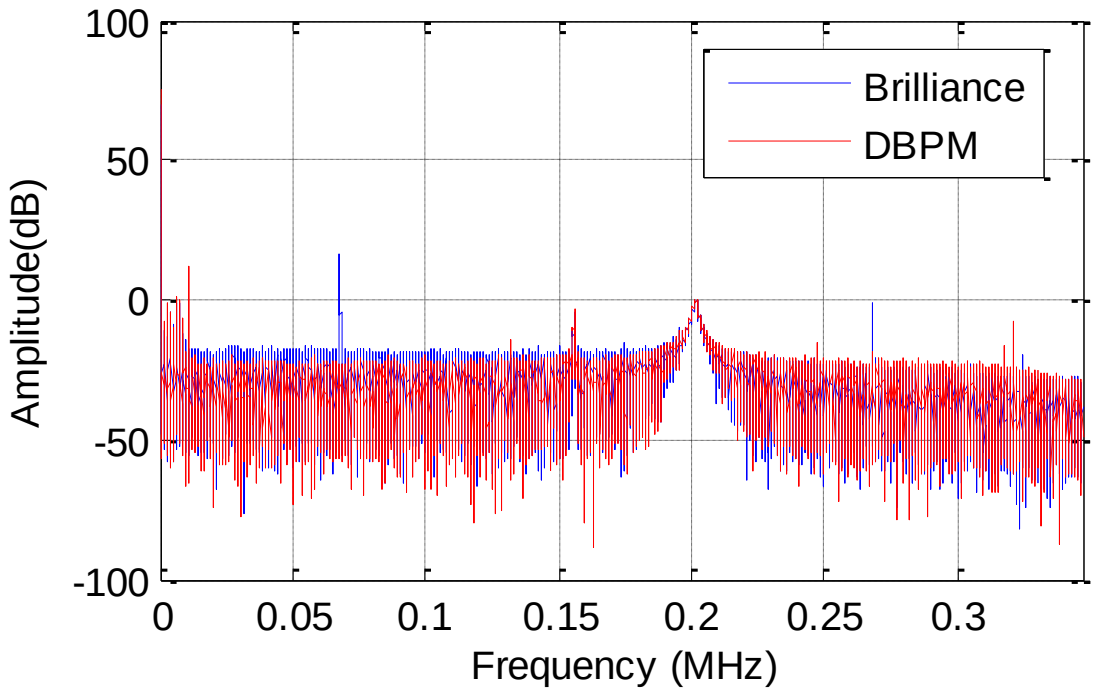

Fig. 17 Frequency Spectrum of the Y position measurement results during injection.

We have finished the design and testing of the electronics systems, and three of them are now used in SSRF. Shown in Fig. 18 is the photograph.

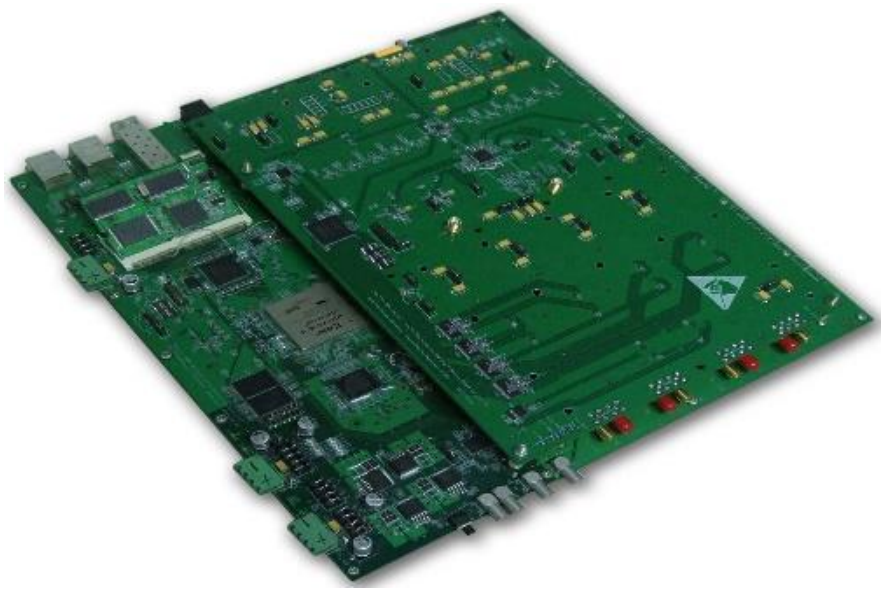

Fig. 18 Photograph of the digital beam position measurement system.

## 4. Digital Beam Phase and Energy Measurement System in the DTL of CSNS

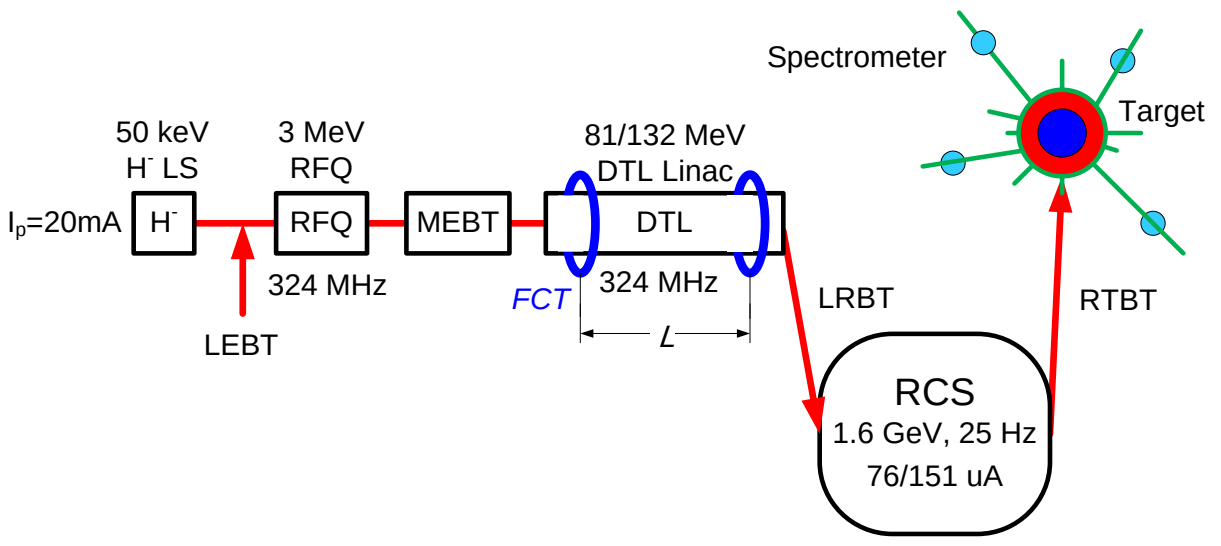

Fig. 19 Layout of CSNS.

As shown in Fig. 19, CSNS consists of $H^-$ ion source, Radio Frequency Quadrupole (RFQ), Drifting Tube Linac (DTL), Rapid Cycling Synchrotron (RCS) and the target. To guarantee the beam quality in DTL, the system we designed import beam RF signals from Fast Current Transformers (FCTs) in the DTL, and then calculate the beam phase and energy for the beam tuning. Based on the time of flight technique [8], the beam energy can be measured with the beam phase difference information between a pair of FCTs with a known distance. Therefore, the kernel task is beam phase measurement.

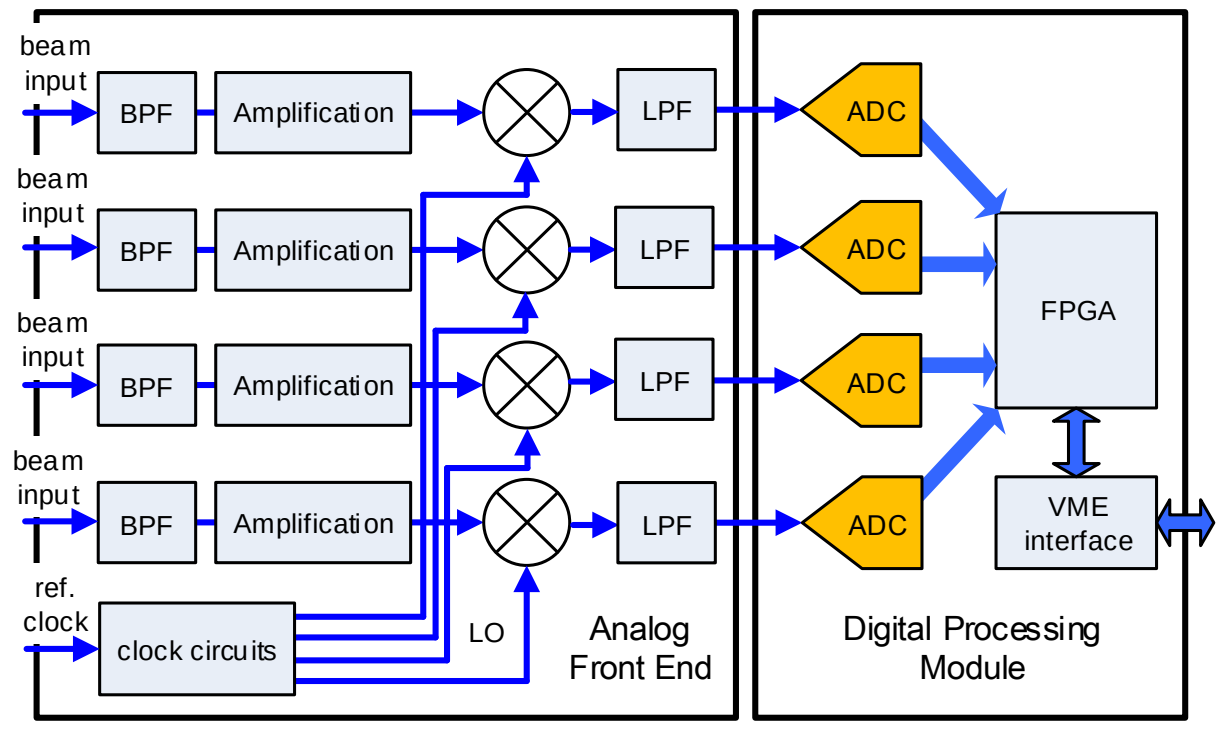

Fig. 20 Block diagram of the beam phase and energy measurement system.

Shown in Fig. 20 is the block diagram of this beam measurement system [24]. The input RF signal is filtered by a BPF, amplified, and then down converted to an IF signal through mixer and LPF. Then this IF signal is digitized with a sampling frequency of four times of the IF signal frequency, through which the I and Q arrays are obtained. Through the DSP algorithms integrated in the FPGA, the beam phase information can be finally obtained.

The input beam signals are pulses with a repetition frequency of more than 300 MHz, and its leading edge is around 200 ps. These pulses are further modulated by a macro pulse (repetition frequency is 25 Hz, pulse width is from 50 μs to 500 μs), and a second modulation is conducted with a fine macro pulse with a repetition frequency of 1 MHz and duty ratio of 40:60 to 80:20. Therefore, the beam signal exhibits a complex frequency spectrum; to confirm the validity of the signal processing method, simulations were conducted [25].

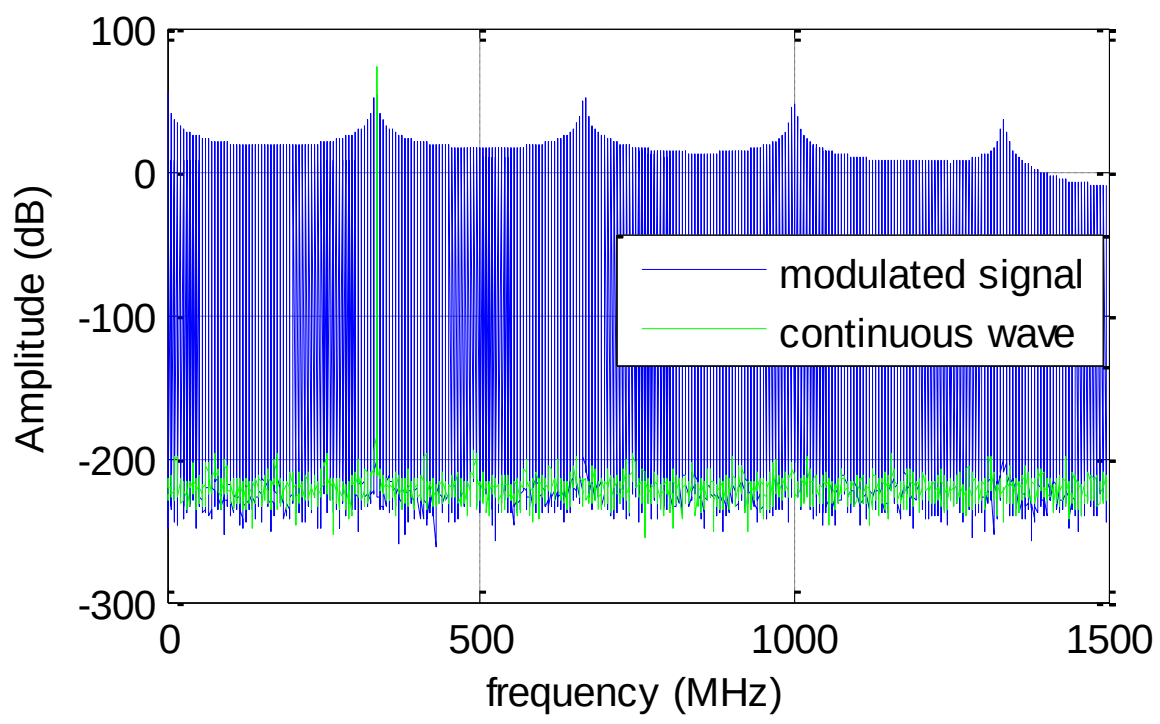

Fig. 21 Frequency spectrum of the modulated signal.

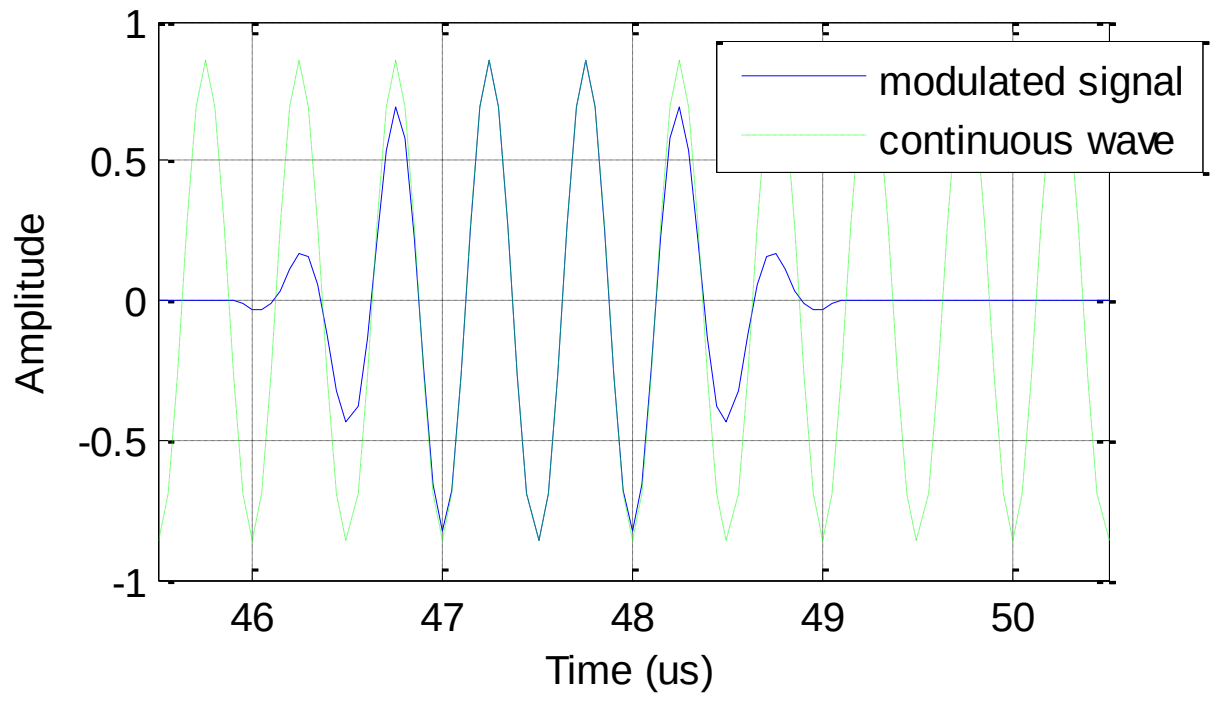

Fig. 22 Simulation results of the signal after down conversion.

As shown in Fig. 21, the frequency is quite complex (as marked in blue color in Fig. 21), as expected. Shown in Fig. 22 is the waveform of the IF signal after down conversion, which indicates almost no distortion in the middle part of the macro pulse despite the waveform at the two ends. The above simulation results verified the correctness of the signal processing method.

Since the IF signal contains exactly four samples in each period after A/D conversion, i.e. $I$, $Q$, $-I$, and $-Q$. Then we can calculate the beam phase based on DSP algorithms, which are integrated within one single FPGA, as shown in Fig. 23.

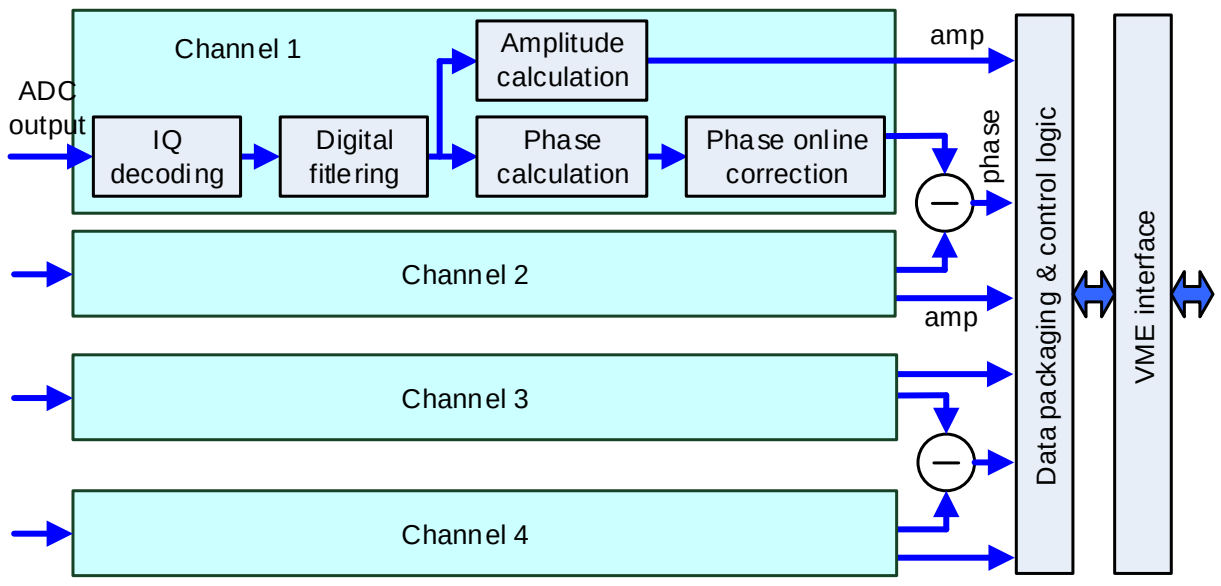

Fig. 23 Block diagram of DSP algorithms integrated in the FPGA.

Shown in Fig. 24 and Fig. 25 are the beam phase and energy measurement system and the test platform.

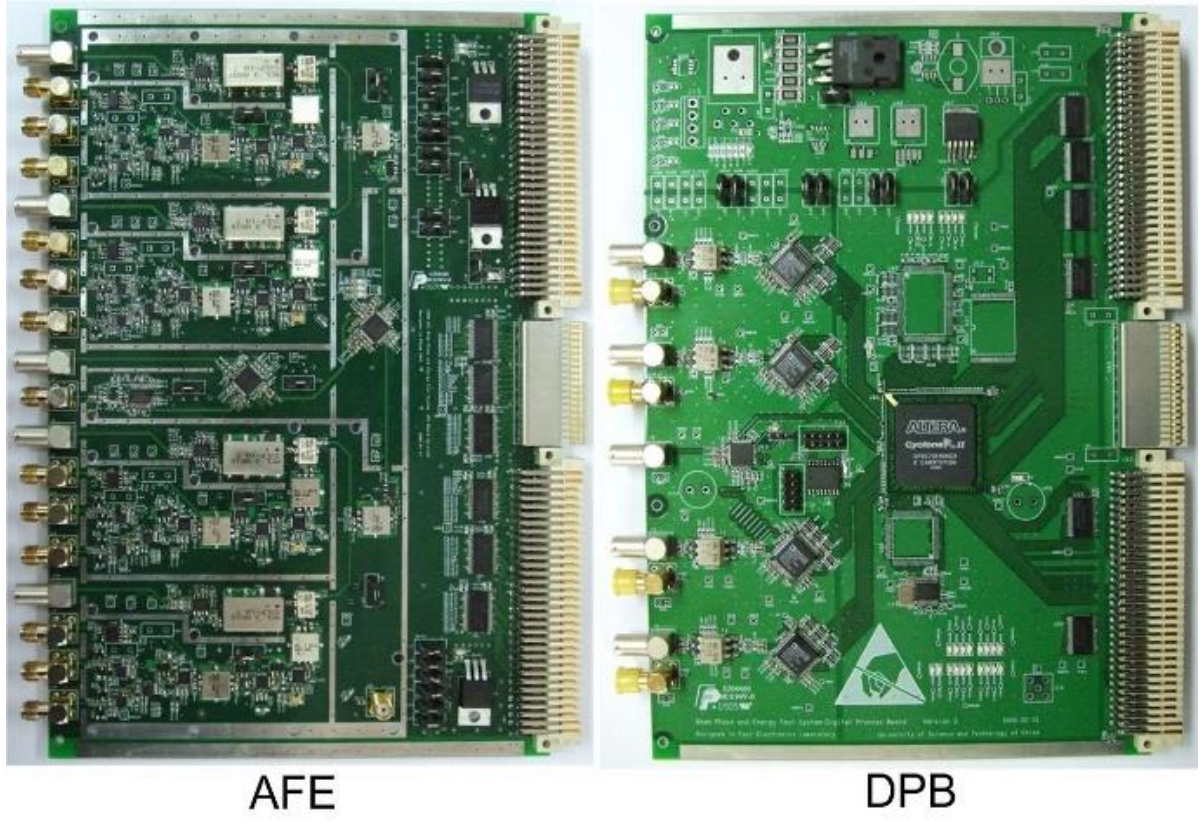

Fig. 24 Photograph of the Analog Front End (AFE) and Digital Processing Board (DPB).

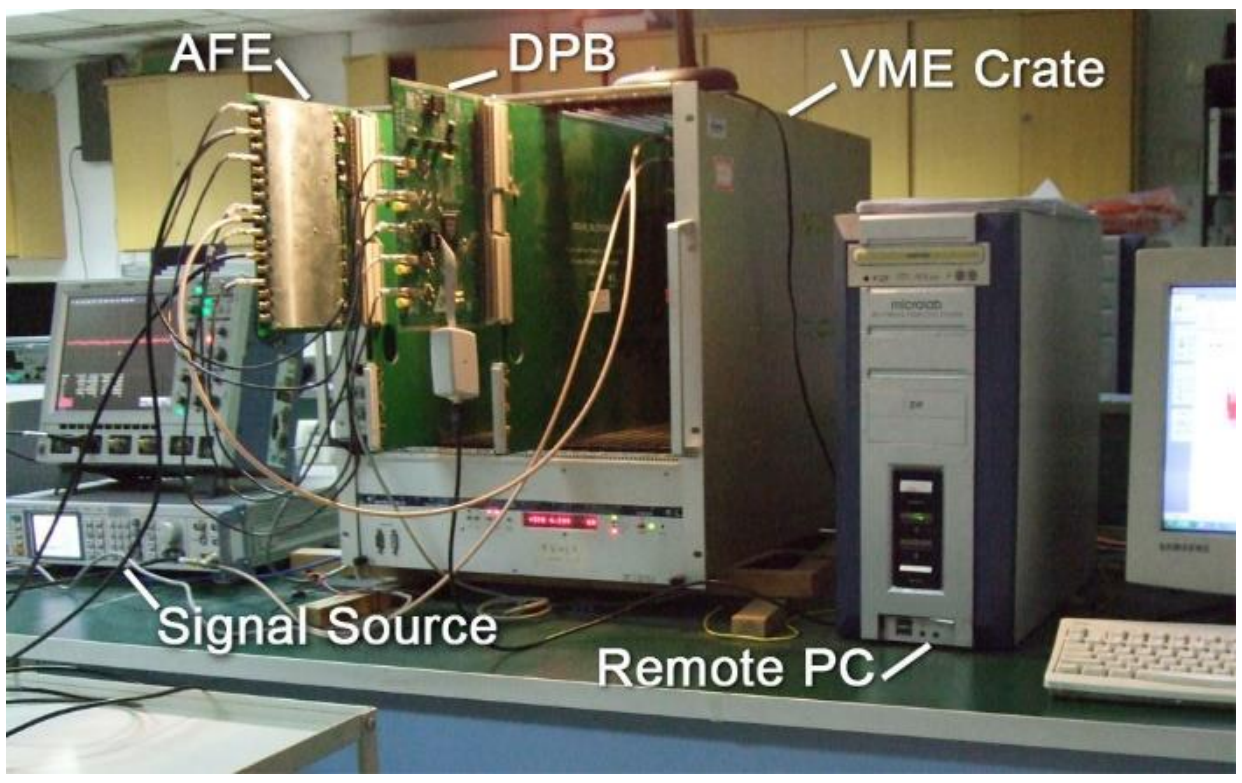

Fig. 25 System under test.

Fig. 26 and Fig. 27 are the test results of the waveform of digitized IF signal, which indicate there exist 4 samples in each period, well as expected.

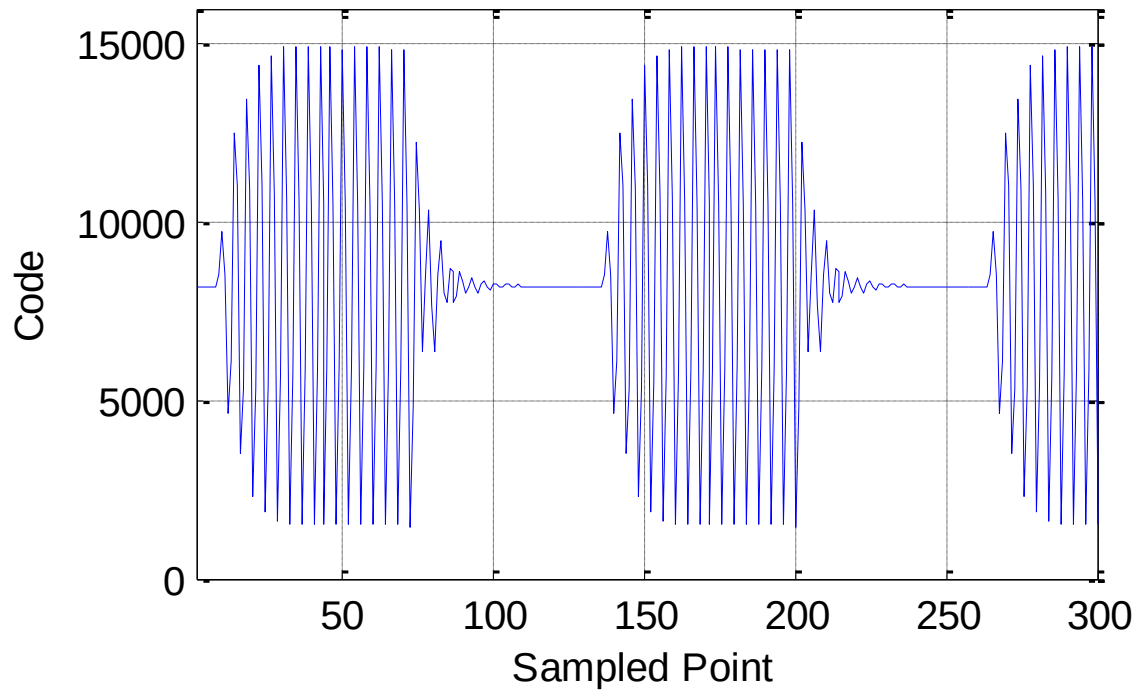

Fig. 26 Waveform of the digitized IF signal.

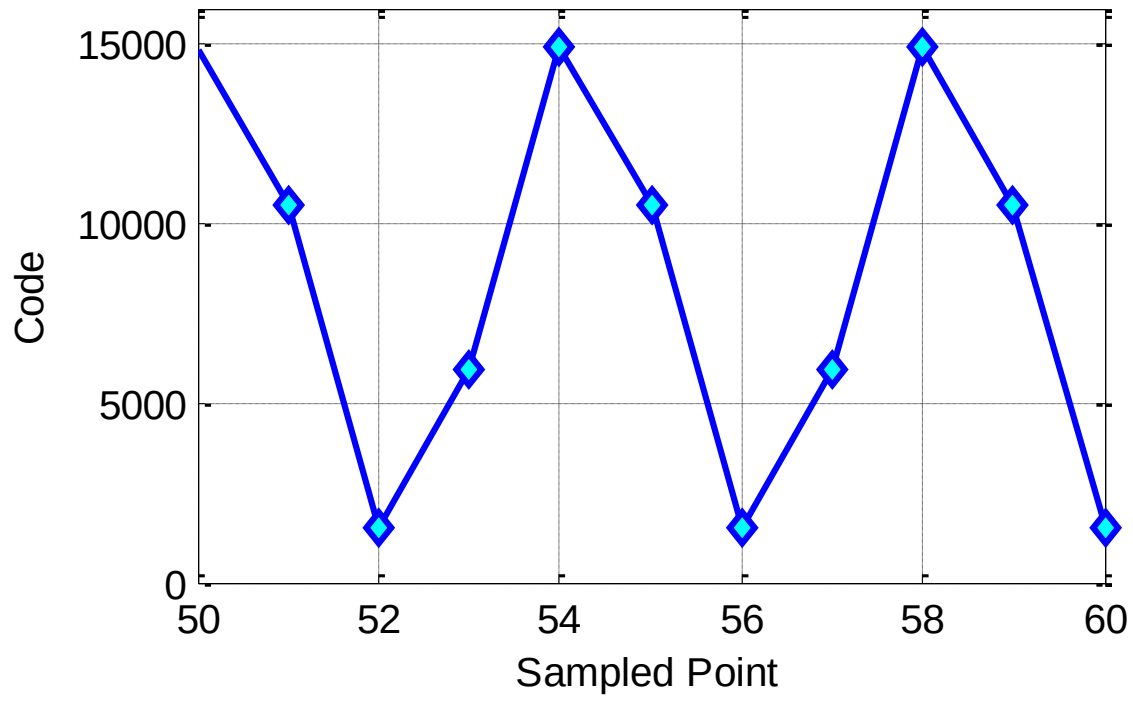

Fig. 27 Detailed waveform of the digitized IF signal.

Fig. 28 is the phase resolution test result. The phase resolution is better than 0.1 °(@ 367 kHz) over a dynamic range from -50 dBm to 7 dBm, well beyond the required ±0.5 °.

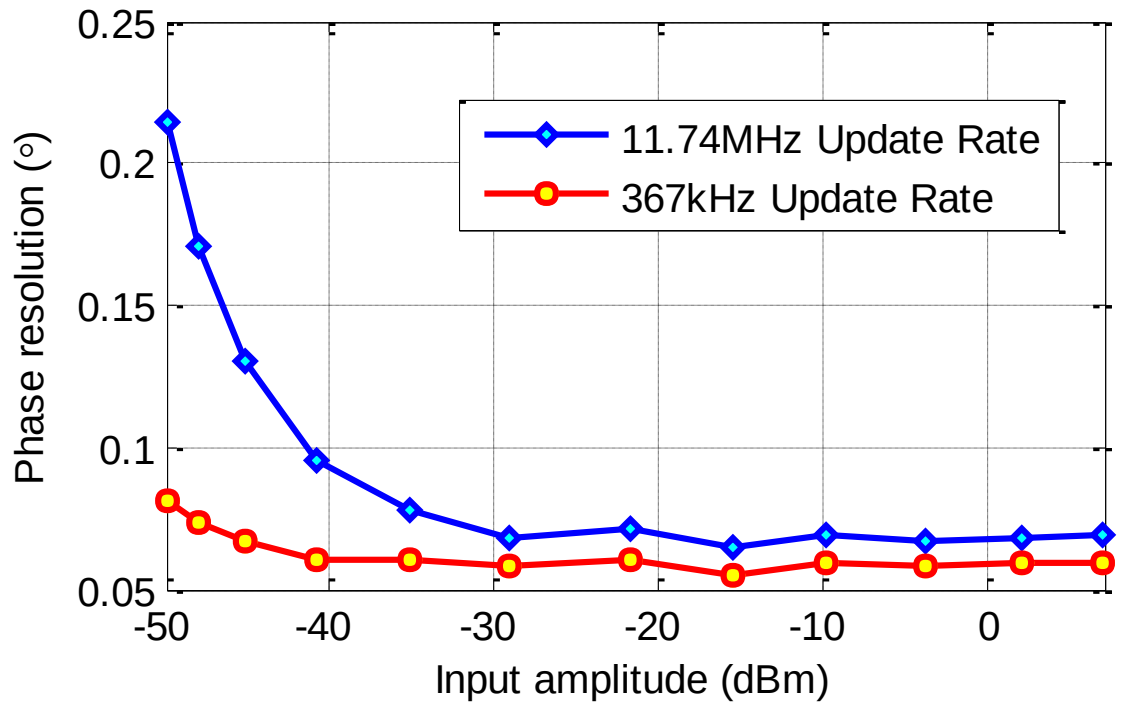

Fig. 28 Phase resolution test results.

## 5. Digital Beam Phase and Position Measurement System in the Proton Linac of ADS

Based on the above research, we proposed a new method to simplify the beam measurement electronics, in which the input beam RF signals are directly under sampled. By precisely tuning the sampling frequency using Phase Locked Loop (PLL) chips, orthogonal streams (i.e. I and Q arrays) can be obtained. The principle of the technique is shown in Fig. 29.

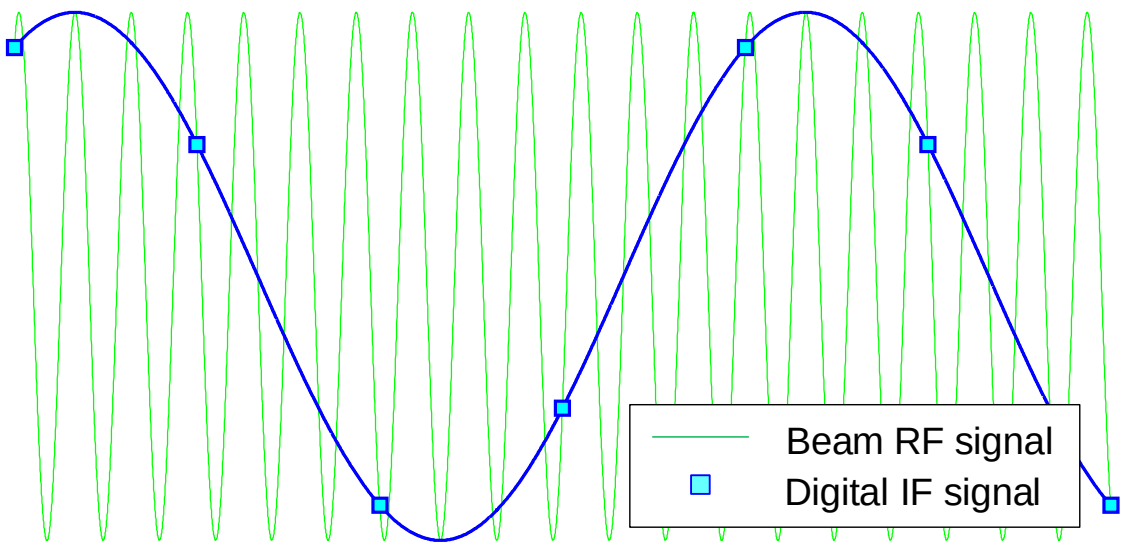

Fig. 29 Principle of the direct IQ under sampling method.

Since the I and Q arrays are obtained directly through A/D conversion, compared with the above two systems, complexity of both the analog circuits and DSP algorithms are greatly reduced. We applied this new method in the beam measurement of the proton Linac in ADS, and integrated both beam phase and position measurement within one single instrument.

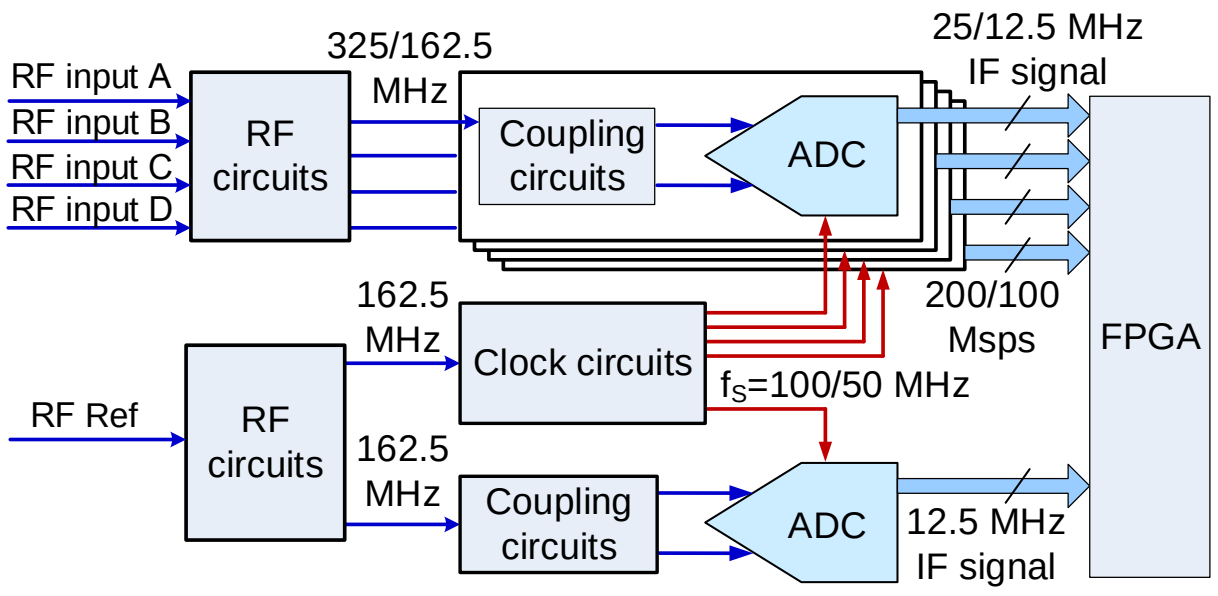

Fig. 30 Block diagram of the beam phase and position measurement system.

As shown in Fig. 30, the input signals from the four pickups of BPM are first amplified and filtered by the RF circuits and then directly under sampled by high-speed high-resolution ADCs [26]. By tuning the sampling frequency ($f_S$) using cascaded PLLs, a certain relationship between $f_S$ and the RF signal frequency ($f_{RF}$) can be guaranteed, as in

$$f_S = \frac{4 f_{RF}}{4M \pm 1} \tag{7}$$

where M is an integer. Through A/D conversion, a digital IF signal with a frequency of $f_S/4$ can be obtained, i.e. there exist four samples ($I$, $Q$, $-I$, $-Q$) within each IF period.

In this system, two schemes are studied; one is based on the signal processing of the fundamental frequency component of 162.5 MHz, i.e. the repetition frequency of the beam signal, and the other is based on 325 MHz, i.e. the second harmonic of the beam signal. The corresponding sampling frequencies for the above two schemes are 50 MHz and 100 MHz, respectively, which result in digital IF signals of 12.5 MHz and 25 MHz, in good concord with (7). Shown in Fig. 31 and Fig. 32 are the electronics modules and the system under test.

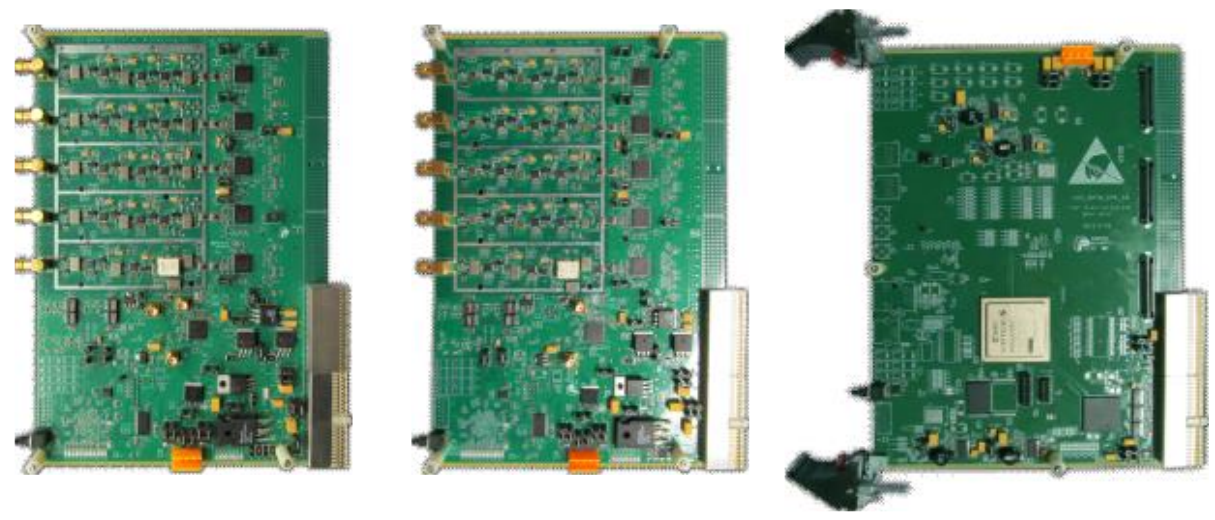

a) AFE 325 MHz b) AFE 325 MHz c) DPB

Fig. 31 Photograph of the Analog Front Ends (AFEs) and Digital Processing Board (DPB).

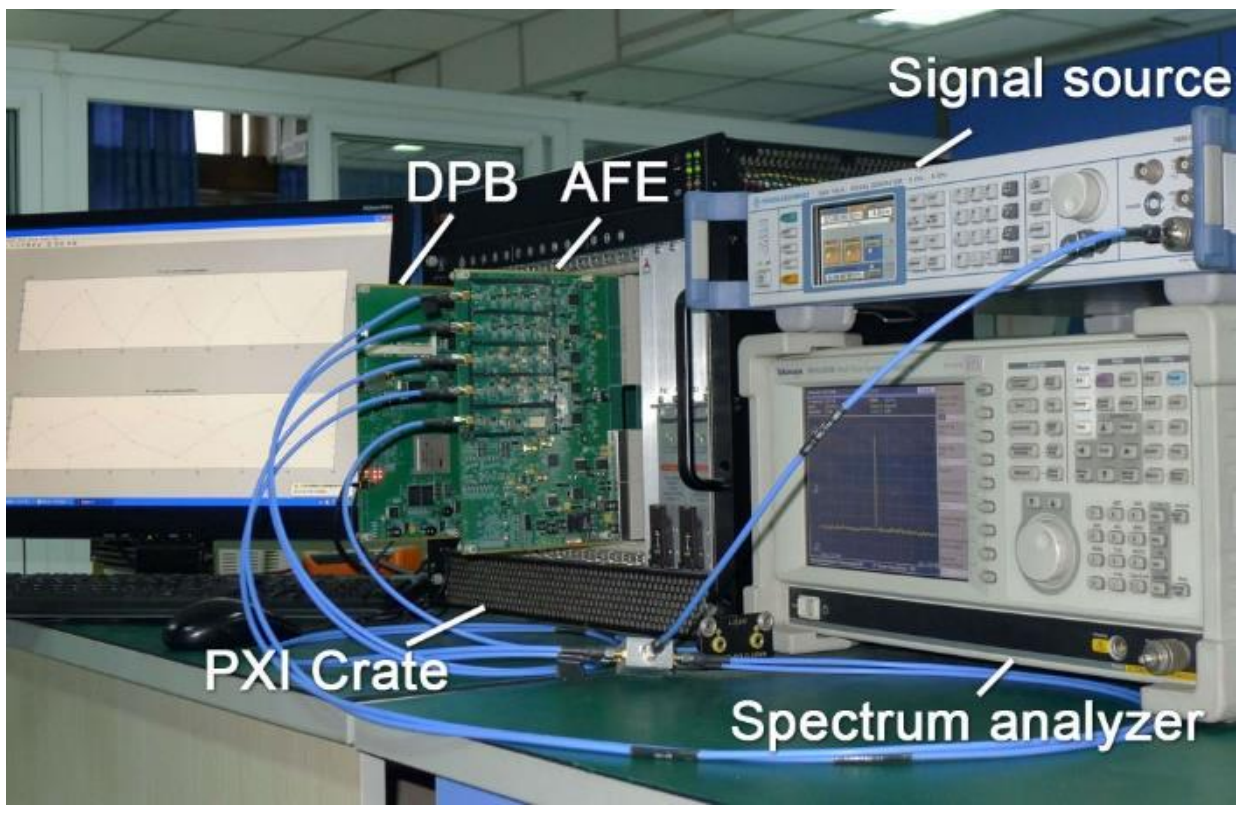

Fig. 32 System under test.

Shown in Fig. 33 and Fig. 34 are the test results of the beam phase and position resolution. As for these two schemes, a phase resolution better than 0.2 °and a position resolution better than 30 μm are both successfully achieved over a dynamic range from -60 dBm to 0 dBm, well beyond the application requirement.

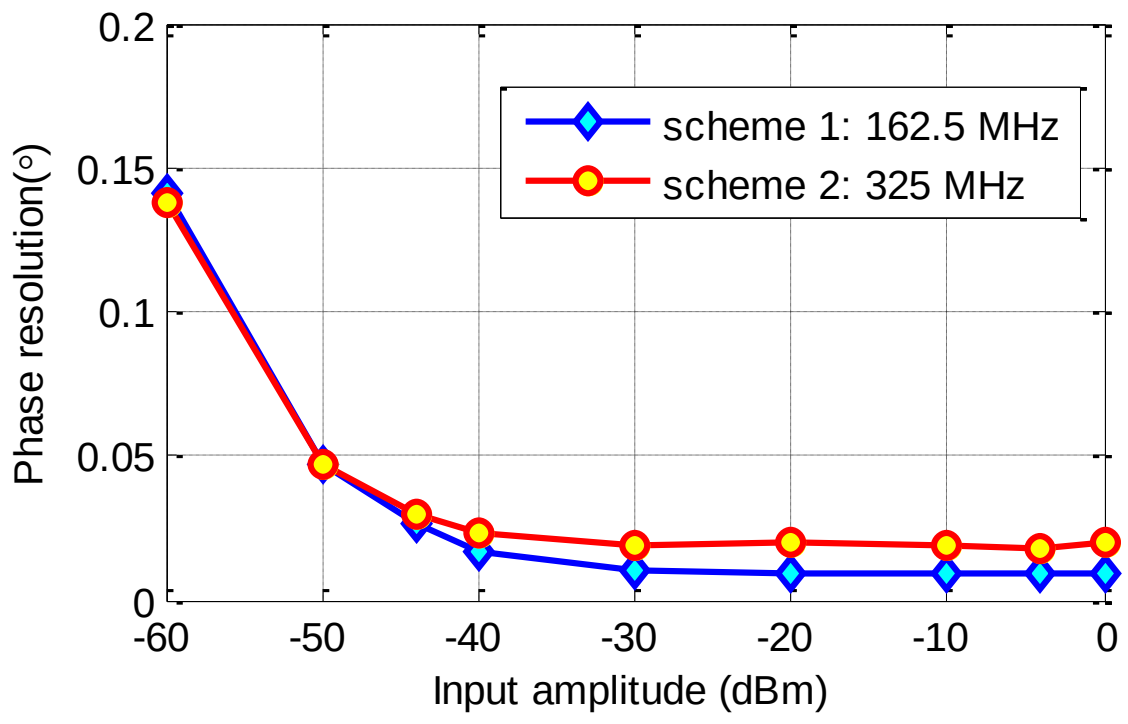

Fig. 33 Phase resolution test results.

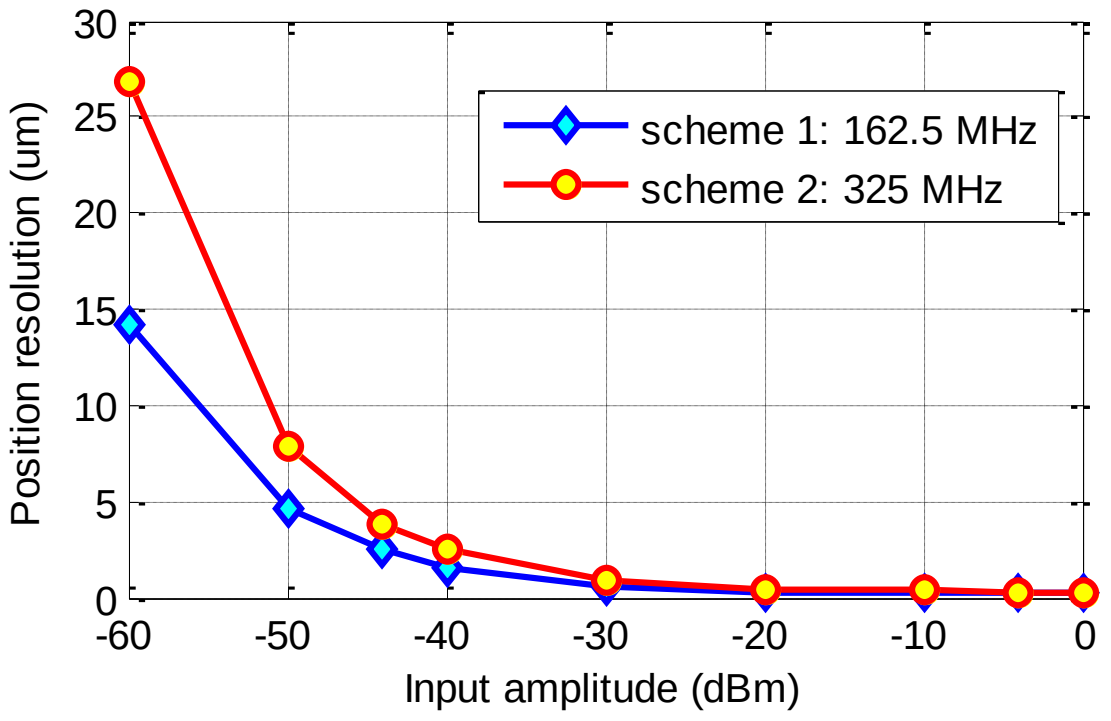

Fig. 34 Position resolution test results.

## 4 Summary

In this paper, we reviewed the basic beam phase and position measurement techniques, discussed the modern digital beam measurement methods based on the IQ analysis. Three beam measurement electronics systems that we designed were also presented, which are applied or planned to be applied in accelerators in China. Based on the above research, we also expect to apply these techniques in beam measurement of future accelerators.

*The authors would like to thank the collaborators in SINAP, IHEP, and IMP, CAS for their help, which makes this work possible.*

---

## Reference

1 J. Hinkson, J. Johnston, I. Ko. Advanced Light Source (ALS) Beam Position Monitor. Proc. Particle Accelerator Conf., 1989, **3**: 1507-1509

2 J. Y. Huang, D. H. Jung, D. T. Kim, et al. PLS Beam Position Monitoring System. 3rd European Particle Accelerator Conference, Berlin, Germany, 1992, pp. 1115-1117

3 W. H. Hwang, H. G. Kim, W. W. Lee, et al. Measurement of Beam Energy Drift in PLS 2.5 GeV LINAC. 3rd Asian Particle Accelerator Conference, Gyeongju, Korea, 2004, pp. 300-302

4 P. Gu, Z. Geng, and G. Pei et al. RF phasing system for BEPCII LINAC. 3rd Asian Particle Accelerator Conference, Gyeongju, Korea, 2004, pp. 288-290

5 Z. Geng, P. Gu, and M. Hou et al. Design and calibration of a phase and amplitude detector. Proc. Particle Accelerator Conf., Knoxville, Tennessee, 2005, pp. 1-3

6 J. Yue L. Ma, J. Cao, L. Wang, Front-end Electronics for BEPCII transverse feedback system. High Power Laser and Particle Beams, 2005, **17**(6): 951-955

7 J. Power and M. Stettler. The Design and Initial Testing of a Beam Phase and Energy Measurement for LEDA. Proceeding of the 1998 Beam Instrumentation Workshop (BIW98) conference

8 Masanori Ikegami, et al. RF Tuning Schemes for J-PARC DTL and SDTL. Proceedings of LINAC 2004, Lübeck, Germany, pp. 414-416

9 J. F. Power, J. D. Gilpatrick, and M. W. Stettler. Design of A VXI Module for Beam Phase and Energy Measurements for LEDA. Proceedings of the Particle Accelerator Conference, 1997, 2: 2041-2043

10 J. Power, J. O'Hara, S. Kurennoy, M. Plum, and M. Stettler. Beam position monitors for the SNS LINAC. Proc. Particle Accelerator Conf., 2001 2: 1375-1377

11 Harald Klingbeil. A Fast DSP-Based Phase-Detector for Closed-Loop RF Control in Synchrotrons. IEEE Trans. Instrum. Meas., 2005, 54: 1209-1213

12 LAI Long-Wei, LENG Yong-Bin, YAN Yin-Bing, et al. The algorithm for digital BPM signal processing. Nuclear Techniques, 2010, **33**(10): 734-739

13 YI Xing, LENG Yong-Bin, LAI Long-Wei, et al. RF front-end for digital beam position monitor signal processor. Nuclear Science and Techniques, 2011, **22**(2): 65-69

14 SUN Bao-Gen, WANG Jun-Hua, LU Ping, et al. Upgrade of beam diagnostics in NSRL Phase II Project. Journal of University of Science and Technology of China, 2007, **37**(4-5): 364-370

15 SUN Bao-Gen, CAO Yong, Li Ji-Hao, et al. Recent research advances of beam diagnostic system for HLS. Journal of University of Science and Technology of China, 2007, **37**(4-5): 475-482

16 YU Lu-Yang, YIN Chong-Xian, LIU De-Kang. Digital RF phase detector for LINAC in FEL accelerator. Nuclear Science and Techniques, 2005, **16**(2): 76-81

17 C.X. Yin, D.K. Liu, L.Y. Yu. Digital Phase Control System For SSRF LINAC. Proceedings of ICALEPCS07, pp.717-719

18 Kosicek A. Libera electron beam position processor. Proceedings of the 2005 Particle Accelerator Conference. IEEE, 2005: 4284-4286

19 Libera Single Pass H specification. Instrumentation Technologies

20 Han Yan, Lei Zhao, Shubin Liu, Kai Chen, Weihao Wu, Qi An, Yongbin Leng, Xing Yi, Yingbing Yan, Longwei Lai. A beam position measurement system of fully digital signal processing at SSRF. Nuclear Science and Techniques, 2012, **23**: 75-82

21 CHEN Kai, LIU Shubin, YAN Han, WU Weihao, ZHAO Lei, AN Qi, LENG Yongbin, YI Xing,YAN Yingbing, LAI Longwei. An embedded single-board computer for BPM of SSRF. Nuclear Science and Techniques, 2011, **22**(4): 193-199

22 Han Yan, Shubin Liu, Kai Chen, Weihao Wu, Lei Zhao, Qi An, Yongbin Leng,Xing Yi, Yingbing Yan, Longwei Lai. The Design and Initial Testing of the Beam Position Measurement System in SSRF Based on Fully Digital Signal Processing. International Workshop on ADC Modeling, Testing and Data Converter Analysis and Design and IEEE 2011 ADC Forum, Orvieto, Italy, June 30th - July 1st, 2011

23 Hao Zhou, Shubin Liu, Kai Chen, Weihao Wu, Lei Zhao, Qi An. Design of the Fully Digital Beam Position Monitor for Beam Position Measurement in SSRF. 9th International Conference on Electronic Measurement and Instruments, 2009, **1**: 1045-1051

24 Lei Zhao, Shubin Liu, Shaochun Tang, Qi An. The Design and Initial Testing of the Beam Phase and Energy Measurement System for DTL in the Proton Accelerator of CSNS. IEEE Transactions on Nuclear Science, 2010, **57**(2): 533-538

25 ZHAO Lei, LIU Shu-Bin, AN Qi, The Research of the Beam Phase and Energy Test System for DTL in the Proton Accelerator of CSNS. Nuclear Electronics & Detection Technology, 2009, **29**(5): 1005-1011

26 Lei Zhao, Xingshun Gao, Xiaofang Hu, Shubin Liu, Qi An. Beam Position and Phase Measurement System for the Proton

Accelerator in ADS. IEEE Transactions on Nuclear Science, 2014, **61**(1): 538-545